\documentclass[aps,twocolumn,prl,showpacs,showkeys,preprintnumbers,superscriptaddress,nobibnotes,floatfix,longbibliography,notitlepage,nofootinbib]{revtex4-2}

\usepackage{amsmath}
\usepackage{amsfonts}
\usepackage{amssymb}
\usepackage{graphicx}
\usepackage[colorlinks=true,allcolors=blue]{hyperref}
\usepackage{relsize}
\usepackage{gensymb}
\usepackage{fontawesome}
\usepackage[capitalize]{cleveref}
\usepackage[dvipsnames,table]{xcolor}

\usepackage{lineno}
\begin{document}

\title{A search for an eV-scale sterile neutrino using improved high-energy $\nu_\mu$ event reconstruction in IceCube}

\affiliation{III. Physikalisches Institut, RWTH Aachen University, D-52056 Aachen, Germany}
\affiliation{Department of Physics, University of Adelaide, Adelaide, 5005, Australia}
\affiliation{Dept. of Physics and Astronomy, University of Alaska Anchorage, 3211 Providence Dr., Anchorage, AK 99508, USA}
\affiliation{Dept. of Physics, University of Texas at Arlington, 502 Yates St., Science Hall Rm 108, Box 19059, Arlington, TX 76019, USA}
\affiliation{CTSPS, Clark-Atlanta University, Atlanta, GA 30314, USA}
\affiliation{School of Physics and Center for Relativistic Astrophysics, Georgia Institute of Technology, Atlanta, GA 30332, USA}
\affiliation{Dept. of Physics, Southern University, Baton Rouge, LA 70813, USA}
\affiliation{Dept. of Physics, University of California, Berkeley, CA 94720, USA}
\affiliation{Lawrence Berkeley National Laboratory, Berkeley, CA 94720, USA}
\affiliation{Institut f{\"u}r Physik, Humboldt-Universit{\"a}t zu Berlin, D-12489 Berlin, Germany}
\affiliation{Fakult{\"a}t f{\"u}r Physik {\&} Astronomie, Ruhr-Universit{\"a}t Bochum, D-44780 Bochum, Germany}
\affiliation{Universit{\'e} Libre de Bruxelles, Science Faculty CP230, B-1050 Brussels, Belgium}
\affiliation{Vrije Universiteit Brussel (VUB), Dienst ELEM, B-1050 Brussels, Belgium}
\affiliation{Department of Physics and Laboratory for Particle Physics and Cosmology, Harvard University, Cambridge, MA 02138, USA}
\affiliation{Dept. of Physics, Massachusetts Institute of Technology, Cambridge, MA 02139, USA}
\affiliation{Dept. of Physics and The International Center for Hadron Astrophysics, Chiba University, Chiba 263-8522, Japan}
\affiliation{Department of Physics, Loyola University Chicago, Chicago, IL 60660, USA}
\affiliation{Dept. of Physics and Astronomy, University of Canterbury, Private Bag 4800, Christchurch, New Zealand}
\affiliation{Dept. of Physics, University of Maryland, College Park, MD 20742, USA}
\affiliation{Dept. of Astronomy, Ohio State University, Columbus, OH 43210, USA}
\affiliation{Dept. of Physics and Center for Cosmology and Astro-Particle Physics, Ohio State University, Columbus, OH 43210, USA}
\affiliation{Niels Bohr Institute, University of Copenhagen, DK-2100 Copenhagen, Denmark}
\affiliation{Dept. of Physics, TU Dortmund University, D-44221 Dortmund, Germany}
\affiliation{Dept. of Physics and Astronomy, Michigan State University, East Lansing, MI 48824, USA}
\affiliation{Dept. of Physics, University of Alberta, Edmonton, Alberta, T6G 2E1, Canada}
\affiliation{Erlangen Centre for Astroparticle Physics, Friedrich-Alexander-Universit{\"a}t Erlangen-N{\"u}rnberg, D-91058 Erlangen, Germany}
\affiliation{Physik-department, Technische Universit{\"a}t M{\"u}nchen, D-85748 Garching, Germany}
\affiliation{D{\'e}partement de physique nucl{\'e}aire et corpusculaire, Universit{\'e} de Gen{\`e}ve, CH-1211 Gen{\`e}ve, Switzerland}
\affiliation{Dept. of Physics and Astronomy, University of Gent, B-9000 Gent, Belgium}
\affiliation{Dept. of Physics and Astronomy, University of California, Irvine, CA 92697, USA}
\affiliation{Karlsruhe Institute of Technology, Institute for Astroparticle Physics, D-76021 Karlsruhe, Germany}
\affiliation{Karlsruhe Institute of Technology, Institute of Experimental Particle Physics, D-76021 Karlsruhe, Germany}
\affiliation{Dept. of Physics, Engineering Physics, and Astronomy, Queen's University, Kingston, ON K7L 3N6, Canada}
\affiliation{Department of Physics {\&} Astronomy, University of Nevada, Las Vegas, NV 89154, USA}
\affiliation{Nevada Center for Astrophysics, University of Nevada, Las Vegas, NV 89154, USA}
\affiliation{Dept. of Physics and Astronomy, University of Kansas, Lawrence, KS 66045, USA}
\affiliation{Centre for Cosmology, Particle Physics and Phenomenology - CP3, Universit{\'e} catholique de Louvain, Louvain-la-Neuve, Belgium}
\affiliation{Department of Physics, Mercer University, Macon, GA 31207-0001, USA}
\affiliation{Dept. of Astronomy, University of Wisconsin{\textemdash}Madison, Madison, WI 53706, USA}
\affiliation{Dept. of Physics and Wisconsin IceCube Particle Astrophysics Center, University of Wisconsin{\textemdash}Madison, Madison, WI 53706, USA}
\affiliation{Institute of Physics, University of Mainz, Staudinger Weg 7, D-55099 Mainz, Germany}
\affiliation{Department of Physics, Marquette University, Milwaukee, WI 53201, USA}
\affiliation{Institut f{\"u}r Kernphysik, Westf{\"a}lische Wilhelms-Universit{\"a}t M{\"u}nster, D-48149 M{\"u}nster, Germany}
\affiliation{Bartol Research Institute and Dept. of Physics and Astronomy, University of Delaware, Newark, DE 19716, USA}
\affiliation{Dept. of Physics, Yale University, New Haven, CT 06520, USA}
\affiliation{Columbia Astrophysics and Nevis Laboratories, Columbia University, New York, NY 10027, USA}
\affiliation{Dept. of Physics, University of Oxford, Parks Road, Oxford OX1 3PU, United Kingdom}
\affiliation{Dipartimento di Fisica e Astronomia Galileo Galilei, Universit{\`a} Degli Studi di Padova, I-35122 Padova PD, Italy}
\affiliation{Dept. of Physics, Drexel University, 3141 Chestnut Street, Philadelphia, PA 19104, USA}
\affiliation{Physics Department, South Dakota School of Mines and Technology, Rapid City, SD 57701, USA}
\affiliation{Dept. of Physics, University of Wisconsin, River Falls, WI 54022, USA}
\affiliation{Dept. of Physics and Astronomy, University of Rochester, Rochester, NY 14627, USA}
\affiliation{Department of Physics and Astronomy, University of Utah, Salt Lake City, UT 84112, USA}
\affiliation{Dept. of Physics, Chung-Ang University, Seoul 06974, Republic of Korea}
\affiliation{Oskar Klein Centre and Dept. of Physics, Stockholm University, SE-10691 Stockholm, Sweden}
\affiliation{Dept. of Physics and Astronomy, Stony Brook University, Stony Brook, NY 11794-3800, USA}
\affiliation{Dept. of Physics, Sungkyunkwan University, Suwon 16419, Republic of Korea}
\affiliation{Institute of Basic Science, Sungkyunkwan University, Suwon 16419, Republic of Korea}
\affiliation{Institute of Physics, Academia Sinica, Taipei, 11529, Taiwan}
\affiliation{Dept. of Physics and Astronomy, University of Alabama, Tuscaloosa, AL 35487, USA}
\affiliation{Dept. of Astronomy and Astrophysics, Pennsylvania State University, University Park, PA 16802, USA}
\affiliation{Dept. of Physics, Pennsylvania State University, University Park, PA 16802, USA}
\affiliation{Dept. of Physics and Astronomy, Uppsala University, Box 516, SE-75120 Uppsala, Sweden}
\affiliation{Dept. of Physics, University of Wuppertal, D-42119 Wuppertal, Germany}
\affiliation{Deutsches Elektronen-Synchrotron DESY, Platanenallee 6, D-15738 Zeuthen, Germany}

\author{R. Abbasi}
\affiliation{Department of Physics, Loyola University Chicago, Chicago, IL 60660, USA}
\author{M. Ackermann}
\affiliation{Deutsches Elektronen-Synchrotron DESY, Platanenallee 6, D-15738 Zeuthen, Germany}
\author{J. Adams}
\affiliation{Dept. of Physics and Astronomy, University of Canterbury, Private Bag 4800, Christchurch, New Zealand}
\author{S. K. Agarwalla}
\thanks{also at Institute of Physics, Sachivalaya Marg, Sainik School Post, Bhubaneswar 751005, India}
\affiliation{Dept. of Physics and Wisconsin IceCube Particle Astrophysics Center, University of Wisconsin{\textemdash}Madison, Madison, WI 53706, USA}
\author{J. A. Aguilar}
\affiliation{Universit{\'e} Libre de Bruxelles, Science Faculty CP230, B-1050 Brussels, Belgium}
\author{M. Ahlers}
\affiliation{Niels Bohr Institute, University of Copenhagen, DK-2100 Copenhagen, Denmark}
\author{J.M. Alameddine}
\affiliation{Dept. of Physics, TU Dortmund University, D-44221 Dortmund, Germany}
\author{N. M. Amin}
\affiliation{Bartol Research Institute and Dept. of Physics and Astronomy, University of Delaware, Newark, DE 19716, USA}
\author{K. Andeen}
\affiliation{Department of Physics, Marquette University, Milwaukee, WI 53201, USA}
\author{C. Arg{\"u}elles}
\affiliation{Department of Physics and Laboratory for Particle Physics and Cosmology, Harvard University, Cambridge, MA 02138, USA}
% ...

\collaboration{IceCube Collaboration}
\thanks{analysis@icecube.wisc.edu}
\noaffiliation

\begin{abstract}
This Letter presents the result of a 3+1 sterile neutrino search using 10.7 years of IceCube data. 
We analyze atmospheric muon neutrinos that traverse the Earth with energies ranging from 0.5 to 100 TeV, incorporating significant improvements in modeling neutrino flux and detector response compared to earlier studies. 
Notably, for the first time, we categorize data into starting and through-going events, distinguishing neutrino interactions with vertices inside or outside the instrumented volume, to improve energy resolution.
The best-fit point for a 3+1 model is found to be at $\sin^2(2\theta_{24}) = 0.16$ and $\Delta m^{2}_{41} = 3.5$ eV$^2$, which agrees with previous iterations of this study.
The result is consistent with the null hypothesis of no sterile neutrinos with a p-value of 3.1\%.
\end{abstract}

\maketitle

\textbf{Introduction.---}Anomalies identified in short-baseline oscillation experiments~\cite{PhysRevLett.77.3082,PhysRevLett.110.161801,PhysRevLett.128.232501} pose a challenge to the established three-mass neutrino framework. 
A minimal explanation for these observed anomalies---the ``3+1" model---postulates the existence of an additional neutrino mass state, $\nu_4$, predominantly consisting of a flavor that does not couple to weak interactions to avoid collider constraints~\cite{ALEPH:2005ab}.
These anomalous results favor a 3+1 scenario with mass-squared differences ranging from 0.1 to 10\,eV$^2$ relative to a model devoid of sterile neutrinos~\cite{Gariazzo:2017fdh,Dentler:2018sju,Diaz:2019fwt,Dasgupta:2021ies,Abdullahi:2023ejc}. 

However, because of flavor mixing, results from $\nu_\mu \rightarrow \nu_e$ appearance, $\nu_e \rightarrow \nu_e$ disappearance, and $\nu_\mu \rightarrow \nu_\mu$ disappearance are linked; thus the anomalies, which are only observed in the $\nu_\mu \rightarrow \nu_e$ and $\nu_e \rightarrow \nu_e$ channels predict oscillation parameters for $\nu_\mu \rightarrow \nu_\mu$.
The non-observation of $\nu_\mu \rightarrow \nu_\mu$ disappearance oscillations~\cite{PhysRevLett.107.011802,PhysRevLett.108.191801,PhysRevD.84.071103,PhysRevLett.117.151803,PhysRevLett.52.1384,PhysRevD.91.052019,PhysRevD.85.032007,PhysRevD.86.052009} in the relevant parameter region represents a very serious challenge to the model.
In this confusing situation, for conclusive statements regarding the 3+1 scenario, it is imperative to further investigate $\nu_\mu$ disappearance with increased sample size and systematic control.
This motivates the analysis we present in this Letter.

\begin{figure}[h!]
    \centering
    \includegraphics[width=0.5\textwidth]{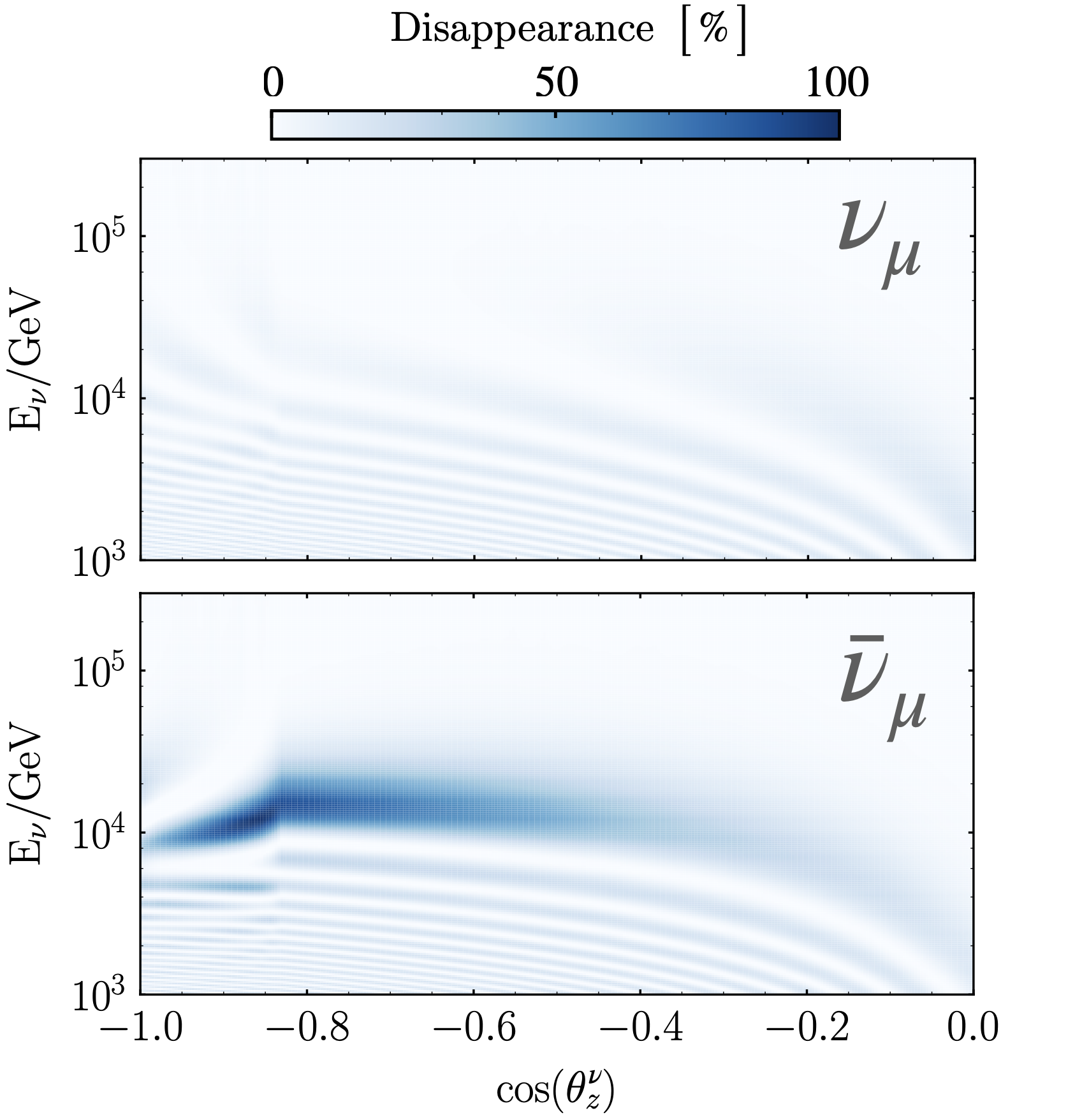}
    \caption{\textbf{Oscillograms}: Muon neutrino (top) and antineutrino (bottom) disappearance probability as a function of the neutrino energy $E_\nu$ and cosine of the zenith angle $\theta_z^\nu$ (proportional to neutrino pathlength) assuming the best-fit sterile neutrino hypothesis of this analysis, i.e., $\sin^2(2\theta_{24}) = 0.16$ and $\Delta m^{2}_{41} = 3.5$ eV$^2$.}
    \label{fig:flux}
\end{figure}

Stringent constraints on sterile-neutrino-induced $\nu_\mu$ disappearance arise from IceCube's study neutrinos produced in the atmosphere with TeV energies, known as high-energy atmospheric neutrinos~\cite{IceCube:2016rnb,IceCube:2020phf,IceCube:2020tka}.
IceCube, a gigaton ice-Cherenkov detector near the geographic South Pole~\cite{IceCube:2016zyt}, detects neutrinos produced in cosmic-ray air showers.
The detector consists of an array of photomultiplier tubes in glass pressure housings called ``Digital Optical Modules'' (DOMs)~\cite{Abbasi:2008aa}, deployed in 86 strings.
In the presence of a sterile neutrino, high-energy atmospheric neutrinos detected by IceCube experience oscillation effects beyond vacuum-like oscillations~\cite{Akhmedov:1988kd,Krastev:1989ix,Chizhov:1998ug,Chizhov:1999az,Akhmedov:1999va} including, for eV-scale masses, matter-enhanced resonant effects induced as high-energy antineutrinos cross the Earth's core~\cite{Nunokawa:2003ep,Choubey:2007ji,Barger:2011rc,Esmaili:2012nz,Esmaili:2013vza,Lindner:2015iaa}. 
Notably, while standard three-neutrino oscillations are negligible at TeV energies due to their mass splitting, the presence of an eV-scale sterile neutrino yields distinct $\nu_\mu$ and $\bar{\nu}_\mu$ disappearance signatures in the atmospheric neutrino flux, as illustrated in Fig.~\ref{fig:flux}. 
A detailed explanation of the different oscillation signatures can be found in Ref.~\cite{PRD}.

The previous search by IceCube for matter-enhanced signatures used seven years of high-energy atmospheric neutrino data~\cite{IceCube:2020phf,IceCube:2020tka}, with over 300,000 up-going\footnote{We refer to events as up-going when they are reconstructed with $\cos\theta_{z}^{\nu}<0$, indicating that they originated below the horizon.} muon tracks in the energy range 0.5 to 10~TeV.
The best-fit point was consistent with the standard
three-neutrino hypothesis at a p-value of 8\%.
In addition, no evidence of sterile-driven $\nu_\mu$ disappearance was observed in a low-energy dataset using the DeepCore subarray~\cite{IceCube:2017ivd,Abbasi:2024ktc}, predominantly composed of sub-100\,GeV neutrinos.

This Letter provides an updated and improved analysis searching for sterile neutrinos using 10.7 years of high-energy neutrino data recorded by IceCube from May 13th, 2011 to June 7th, 2022.
It sets a limit on $\theta_{24}$ compatible with DeepCore's results and the previous high-energy search.
We implement a new event selection that yields a data sample that is larger and higher purity than what was used in previous analyses.
After implementing this new selection, the dataset consists of 368,071 up-going $\nu_\mu$ events with reconstructed energies ranging from 0.5 to 100\,TeV. 
We separate events with vertices inside or outside the instrumented volume, referred to as ``starting" and ``through-going events". 
This differentiation, coupled with a new energy estimator, enables a more precise characterization of the detector response to $\nu_\mu$ interactions, which are substantively different for starting and through-going events. 
The extended energy range of this analysis requires a significant revision of the systematic uncertainty treatment concerning atmospheric and astrophysical neutrino fluxes. 
The differentiation between starting and through-going events also introduces a new approach to evaluating systematic uncertainties associated with the glacial ice.

An in-depth exploration of the technical aspects of the analysis can be found in a complementary article~\cite{PRD}.

\textbf{Event selection and reconstruction.---}At TeV energies, a muon produced in a $\nu_\mu$\footnote{Here, $\nu_x$ refers to either neutrinos or antineutrinos of $x$ flavor, with a similar language for their charged-lepton counterparts} charged-current (CC) interaction can travel several kilometers in ice~\cite{Lipari:1993hd,Koehne:2013gpa}, emitting Cherenkov radiation and producing an elongated light pattern, referred to as track-like signature, as it traverses the instrumented volume.
While a similar pattern is generated by muons originating from cosmic rays in the atmosphere, the earth’s shielding prevents up-going cosmic ray muons from contaminating this sample.
Consequently, we can effectively filter out the majority of atmospheric muons by exclusively considering up-going tracks. 
The track direction is determined with a maximum likelihood method which uses the arrival time distribution of Cherenkov photons registered by the DOMs~\cite{AMANDA:2003vtt,Aartsen:2013vja}. 

The remaining contamination arises from two sources of misreconstructed events: atmospheric muons traversing a limited portion of the detector; and cascade-like events induced by $\nu_{e,\tau}$ CC and all neutral-current (NC) interactions.
We employed a boosted decision tree (BDT)~\cite{Roe:2004na} algorithm to discard such events.
The BDT was trained to identify $\nu_\mu$ CC events using distributions of track-related features, including reconstructed track length, zenith angle, energy, position, morphological track/cascade classifiers, deposited charge in the detector, and a Bayesian likelihood ratio between up-going and down-going track hypotheses~\cite{Weaver:2015bja}. The last was found to be the most used feature by the BDT.
By cutting out events below a specified BDT score, we can eliminate nearly all atmospheric muons and cascade-like events induced by $\nu_{e,\tau}$ CC and $\nu$ NC interactions, resulting in a sample with 99.9\% purity of $\nu_\mu$ CC events.
The remaining 0.1\% background arises mostly from $\nu_\tau$ CC astrophysical events followed by subsequent $\tau\rightarrow\mu$ decay, an irreducible background as they also exhibit a track-like pattern.
Using a BDT increases the signal efficiency by almost a factor of two and reduces atmospheric muon background by half with respect to the previous analysis~\cite{IceCube:2020tka}.

The energy of each event is reconstructed using a convolutional neural network (CNN)~\cite{Abbasi:2021ryj}.
The network is trained with $\nu_\mu$ CC simulated events to estimate the energy deposited by muons and hadronic or electromagnetic showers within the instrumented volume using summary variables that characterize the light observed in each DOM (i.e., overall charge, width of the pulse, time of first hit, etc.).
The energy estimator used in our previous analysis~\cite{IceCube:2020tka} was optimized to infer the deposited energy using information about light deposition from muon tracks~\cite{IceCube:2013dkx}, whereas here, the network is trained to reconstruct the overall energy deposition in the detector.
Therefore, while both approaches perform similarly well in reconstructing the energy for through-going events, the network significantly outperforms the previous estimator in the case of starting events (for which both the hadronic shower and a muon contribute to the energy deposition in the detector).

Given the different performances of the reconstruction for through-going and starting events, a classifier has been devised to differentiate between the two morphologies~\cite{Kronmueller:2019jzh}.
Like the energy estimator, this classifier is also based on a CNN that uses the same pulse-related features from the light observed in each DOM. 
The network is trained with neutrino simulations to assign a score for different morphologies.
By applying a threshold on the score, the sample can be partitioned, resulting in 75\% (25\%) of the events being classified as through-going (starting), with 13\% (0.2\%) of misclassified events in each sample. 

\begin{figure}[h!]
    \centering
    \includegraphics[width=0.5\textwidth]{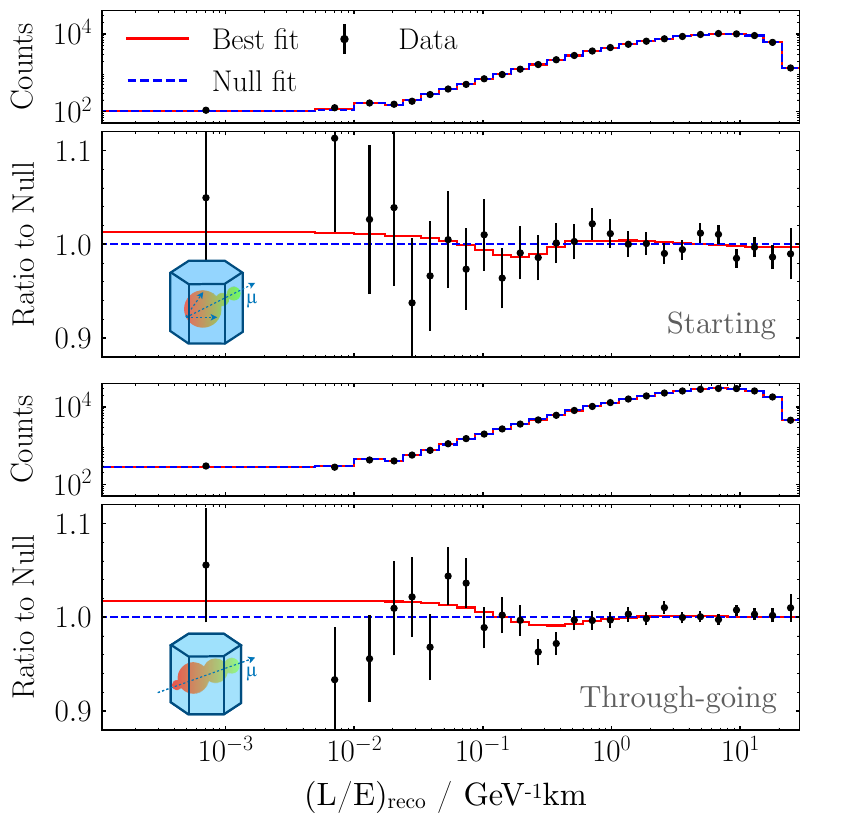}
    \caption{\textbf{$L/E$ distributions}: Data points are black markers with error bars representing the Poissonian statistical error. The solid red and blue lines show the best-fit sterile neutrino hypothesis and the null (no sterile neutrino) hypothesis, respectively, with nuisance parameters set to their best-fit values in each case. The top (bottom) panels show the number of events and the ratio to the null hypothesis in each bin for starting (through-going) events.}
    \label{fig:le}
\end{figure}

The reconstructed $L/E$ distributions for the selected events are shown in Fig.~\ref{fig:le}, showing a good agreement between data and the prediction at the best-fit point of the analysis in both samples.
Different types of events populate different regions of the $L/E$ distribution. 
Horizontal events predominantly cluster at low $L/E$ values, while vertically up-going events are more common in the high $L/E$ region. 
The best-fit expectation for $L/E$ shows a dip at 0.3 km/GeV, primarily due to vacuum oscillations. 
The predicted excess at low $L/E$ arises because the best-fit expectation prefers a higher normalization compared to the null fit, due to fast oscillations happening at $L/E$ below the oscillation maximum. 
This excess is canceled out by the resonant disappearance effects at $L/E$~1km/GeV.
For a lower mass splitting than the best-fit point, the oscillation maximum shifts to larger $L/E$ values. Conversely, the disappearance effect moves toward lower $L/E$ values if $\Delta m^{2}_{41}$ is large, reaching a point where the sample would only be sensitive to the fast oscillations.

\textbf{Analysis.---}Starting and through-going events are binned separately based on their reconstructed energy and the cosine of the zenith angle, $E_{\mathrm{reco}}$ and $\cos\theta_{z}^{\mathrm{reco}}$, respectively. 
Both datasets use the following binning scheme: 24 bins spanning from 0.5 to 100 TeV in $\log_{10}(E_{\mathrm{reco}})$ and 20 bins spanning from -1 to 0 in $\cos\theta_{z}^{\mathrm{reco}}$. 
Each bin's expected and observed event counts are compared using 
the effective likelihood~\cite{Arguelles:2019izp}, which is a modified version of the Poisson likelihood to accommodate Monte Carlo statistical uncertainty. 
Systematic errors are included by weighting the Monte Carlo expectations following the nuisance parameters.
A total of 36 parameters are used in this analysis, as detailed below, for which penalty terms are included in the likelihood based on their prior.
The final likelihood is given by the product of the effective likelihood and the systematic prior probability, $\mathcal{L}=\prod_{i=1}^{N_{bins}}\mathcal{L}_{\mathrm{eff}}(\mu_{i}(\Vec{\theta},\Vec{\eta});n_i)\prod_{j=1}^{N_{syst}}G(\eta_{j})$,
where $n_{i}$ and $\mu_{i}(\Vec{\theta},\Vec{\eta})$ are the observed and expected number of events in the i-th bin; and $\Vec{\theta}$, $\Vec{\eta}$ are the set of physics and nuisance parameters.

Using this binned likelihood function, we conduct a frequentist analysis to test for evidence of eV-scale sterile neutrinos. 
In this study, the sterile neutrino model assumes non-zero $\Delta m_{41}^{2}$ and $\sin^{2}(\theta_{24})$, varying from $10^{-2}$ to $10^{2}$ eV$^2$ and from $10^{-3}$ to 1, respectively. 
The confidence levels are constructed using Wilks' theorem~\cite{10.1214/aoms/1177732360}.
In addition to this approach, we complement our study with a Bayesian analysis, the details of which are presented in~\cite{PRD}.

The expected number of events is computed by convolving the simulated detector response with the expected neutrino flux at the detector. 
The neutrino flux at the Earth's surface is calculated using \texttt{DAEMONFLUX}~\cite{Yanez:2023lsy} for neutrinos generated in the atmosphere. 
The astrophysical neutrino flux is assumed to have an equal flavor composition and be symmetric for neutrinos and antineutrinos~\cite{Bustamante:2015waa,Arguelles:2015dca,Palladino:2015vna,Arguelles:2019tum}; it has an isotropic angular spectrum and an energy spectrum consistent with previous IceCube measurements~\cite{IceCube:2020acn,IceCube:2020wum,IceCube:2021uhz}. 
All these fluxes are propagated to the detector using nuSQuIDS~\cite{Arguelles:2021twb}, assuming the PREM model (using 200 radial
layers)~\cite{DZIEWONSKI1981297} and CSMS cross sections~\cite{Cooper-Sarkar:2011jtt} to model the Earth's density profile and neutrino cross sections, respectively. 
This propagation is conducted independently for each choice of sterile neutrino parameters.

\textbf{Systematic errors.---}The dominant sources of systematic uncertainties for this analysis can be categorized into six groups: conventional and non-conventional neutrino fluxes, normalization, bulk ice properties, local response of the DOMs, and neutrino attenuation.
We quantified the influence of each uncertainty source on the expected sensitivity for this analysis.
At high $\Delta m^2_{41}$, the most significant impact stems from normalization, followed by non-conventional and cosmic-ray flux uncertainties. 
Conversely, at low $\Delta m^2_{41}$, the dominant factors are uncertainties in the ice modeling and hadronic yields.

\textit{Conventional neutrino flux}: 
The uncertainties on the pion and kaon production in the atmosphere are derived using the \texttt{DAEMONFLUX} model.
This model employs \texttt{MCEq}~\cite{Fedynitch:2018cbl} to generate lepton fluxes based on the cosmic-ray spectrum and the hadron yields derived from the Global Spline Fit (GSF~\cite{Dembinski:2017zsh}) and the Data-Driven Hadronic Interaction Model (DDM~\cite{Fedynitch:2022vty}), respectively.
The model provides 6 parameters that control the spectrum of the cosmic rays, the most important being a spectral index and 18 parameters that describe the hadronic yields.
These parameters are calibrated using surface muon flux data. 
The variation of the neutrino flux when modifying each parameter is calculated using the gradient method~\cite{Barr:2006it}.
Parameters linked to low-energy hadron yields have minimal impact within the TeV energy range and thus were not considered as systematics in our analysis.
The model's covariance matrix is included as a prior in our likelihood function, penalizing deviations from the nominal value of the remaining parameters and naturally accounting for their correlations.

The modeling of the conventional flux and its associated uncertainties differ from the previous analysis\footnote{The hadronic interactions and cosmic-ray spectrum were modeled with Sibyll2.3c~\cite{Fedynitch:2018cbl} and HillasGaisser H3a~\cite{Gaisser:2013bla} in the previous analysis~\cite{IceCube:2020tka}. Systematic uncertainties were accounted for following the Barr prescription~\cite{Barr:2006it}.}.
The most notable disparity is evident in the $\nu_\mu/\bar{\nu}_{\mu}$ ratio and its associated error, where \texttt{DAEMONFLUX} confines it to $<10\%$ across the entire energy spectrum. 
For the flux, the error remains below 10\% up to 1 TeV, increasing to 30\% at higher energies.
More details about the new parameterization can be found in Ref.~\cite{PRD}.

The atmospheric density profile and kaon energy losses are also accounted for in assessing conventional neutrino flux uncertainties, following the method outlined in the previous analysis~\cite{IceCube:2020tka}.

\textit{Non-conventional neutrino flux}:
With the maximum reconstructed energy extended from 10 to 100\,TeV compared to the previous analysis~\cite{IceCube:2020tka}, a more conservative approach has been adopted for the high-energy flux, consisting of astrophysical and prompt components. 
This analysis represents these fluxes by a broken power law, with uncorrelated Gaussian priors assigned to the normalization and two spectral indices.
The prior for the normalization at 100 TeV and both spectral indices are centered at  $0.787\times10^{-18}$~GeV$^{-1}$sr$^{-1}$s$^{-1}$cm$^{-2}$ and $-2.5$, respectively, and the width of the priors is 46\% and 36\%.
A uniform prior is assumed for the energy pivot point.
These priors were chosen to encompass all IceCube astrophysical neutrino flux measurements to date.

Mismodeling tests incorporated a galactic component~\cite{IceCube:2023ame}, increasing the prompt component by an order of magnitude or removing it altogether, and considering $\nu$- and $\bar{\nu}$-only astrophysical contributions.
Our investigations revealed a significance for spurious sterile-like signals below 0.3$\sigma$ in all test cases.

\textit{Normalization}: 
An overall normalization term is incorporated in our analysis to model the impact of muon energy loss and neutrino cross section in water/rock on the effective volume and rate of interactions near the detector.
The associated uncertainty of these processes~\cite{Sandrock_2020,Klein:2020nuk,Garcia:2020jwr,Xie:2023suk,Jeong:2023hwe,Candido:2023utz} have a $\mathcal{O}(10\%)$ impact in the scaling factor. 
Thus, a Gaussian prior with a conservative 20\% uncertainty is assigned to this term.
The uncertainties on the ice density and the rock/ice transition region near the detector will also impact the normalization but are negligible (i.e., $<1\%$) compared to those related to the cross section.
The modeling of final state radiation~\cite{Plestid:2024bva}, which changes the fraction of the energy from the neutrino carried out by the outgoing lepton, has been assessed to have a negligible effect on the analysis.

\textit{Bulk Ice Model}:
The presence of impurities between IceCube strings, known as ``dust," affects the scattering and absorption of light within ice~\cite{IceCube:2013llx}.
To propagate the uncertainty on these parameters, we simulate events assuming different dust concentrations using the SnowStorm method~\cite{IceCube:2019lxi}.
The impact of the Fourier coefficients modes was evaluated within the energy-zenith reconstructed space for starting and through-going events. 
The first 5 amplitudes and 4 phases were found to have a significant impact. 
The analysis incorporates nuisance parameters for these modes with correlated Gaussian priors.

\textit{DOM Response and Local Ice Effects}: 
The properties of the ice surrounding the DOMs can affect both the global efficiency and angular dependence of photon detection. 
Following the same approach as the prior analysis, two parameters have been included to model these effects in the reconstructed energy-zenith space for both starting and through-going events. Both parameters are incorporated with an effectively flat prior and allowed to vary
within conservative ranges.

\textit{Neutrino Attenuation}: 
Neutrinos with energies above a few tens of TeV are absorbed as they propagate through the Earth~\cite{Gandhi:1995tf,IceCube:2017roe,IceCube:2020rnc}. 
The uncertainty in this process stems from our imperfect knowledge of the neutrino-nucleus cross section~\cite{Klein:2020nuk,Garcia:2020jwr,Xie:2023suk,Jeong:2023hwe,Candido:2023utz} and the Earth’s composition.
To account for this, scale parameters in the $\nu$ and $\bar{\nu}$ cross sections that alter the Earth absorption~\cite{Vincent:2017svp} have been incorporated.
Gaussian priors with 10\% width are used to encompass the uncertainties in our energy range. Importantly, this parameter is treated independently of the overall normalization, as a distinct impact of nuclear effects is anticipated in light and heavier targets.

\textbf{Results.---}The frequentist analysis found the best-fit point at $\sin^2(2\theta_{24}) = 0.16$ and $\Delta m^{2}_{41} = 3.5$ eV$^2$. 
Compared to the standard three-neutrino hypothesis, the test statistic (TS) is $-2\Delta \log \mathcal{L}=6.96$, yielding a p-value of 3.1\% under the assumption of two degrees of freedom. This probability, corresponding to a significance of 2.2$\sigma$, does not constitute evidence for the existence of an eV-scale sterile neutrino.

To avoid reliance on Wilk's theorem, the p-value was alternatively derived using the Feldman-Cousins procedure~\cite{Feldman:1997qc}. We fitted 200 pseudoexperiments generated at the null point, resulting in p$=$3\%.
Moreover, a Bayesian analysis has been conducted (further details available in~\cite{PRD}), and its results align with those obtained through the frequentist approach.

The data pulls relative to the best fit are normally distributed for starting and through-going events.
Furthermore, we do not find large deviations in the pulls using different bin sizes. 
The goodness-of-fit was extracted by comparing the observed best-fit likelihood with the likelihood distribution of 500 fits to pseudoexperiments generated assuming the best-fit hypothesis, yielding a p-value of 12\%.

\begin{figure}
    \centering
    \includegraphics[width=0.5\textwidth]{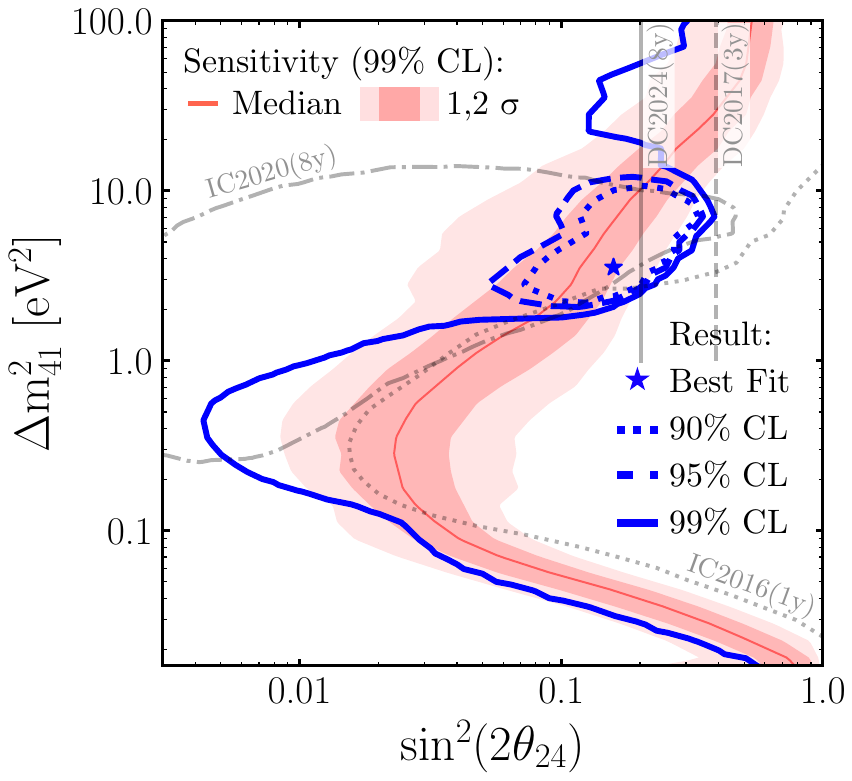}
    \caption{\textbf{Frequentist analysis}. The 90\%, 95\%, and 99\% C.L. contours, assuming Wilks’ theorem, are shown as dotted, dashed, and solid blue lines, respectively. The red bands show the region where 68\% and 95\% of the pseudoexperiment 99\% C.L. observations lie; the red line corresponds to the median. Previous measurements from IceCube~\cite{IceCube:2016rnb,IceCube:2020phf,IceCube:2020tka,IceCube:2017ivd,Abbasi:2024ktc} at 90\% C.L. are shown in grey. }
    \label{fig:result}
\end{figure}

\begin{figure}
    \centering
    \includegraphics[width=0.5\textwidth]{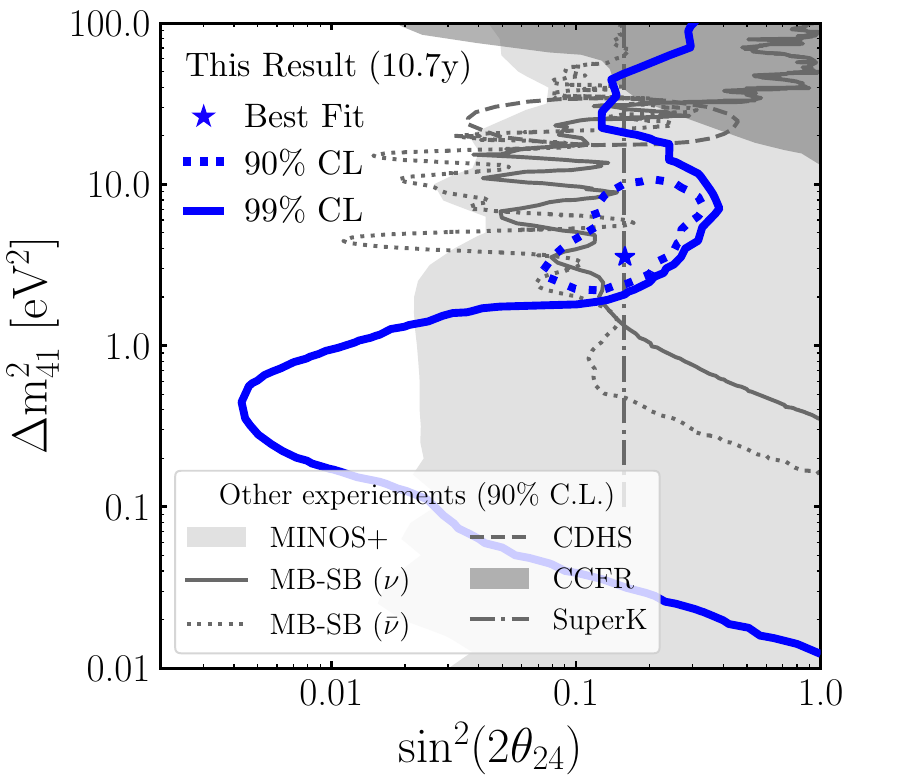}
    \caption{\textbf{Comparison to other experiments}. The 90\% and 99\% C.L. contours (blue lines) of this analysis compared to 90\% C.L. contours from MiniBooNE-SciBooNE~\cite{SciBooNE:2011qyf,MiniBooNE:2012meu}, MINOS~\cite{MINOS:2017cae}, CDHS, CCFR~\cite{Diaz:2019fwt}, and SuperK~\cite{Super-Kamiokande:2014ndf}.}
    \label{fig:comparison}
\end{figure}

Figure~\ref{fig:result} shows the 90\%, 95\%, and 99\% C.L. contours calculated according to Wilks' theorem. 
The sensitivity of this analysis was derived using 500 pseudo-experiments, each of which was generated assuming no sterile neutrino model with the nuisance parameters at their central values. 
The 90\% C.L. preferred region of this analysis is consistent with previous results from IceCube~\cite{IceCube:2016rnb,IceCube:2020phf,IceCube:2017ivd}.
This result is compared with previous measurements of $\nu_\mu$ disappearance from other experiments in Fig.~\ref{fig:comparison}. The 90\% C.L. allowed region from this result indicates an increased tension with the constraints from long-baseline experiments.

The fitted values for the nuisance parameters have very similar behavior for the best-fit sterile and null hypotheses.
The nuisance parameter with the largest difference between null and best-fit hypotheses is the overall normalization, pulling $-0.05\sigma$ and $0.41\sigma$, respectively.
None of the parameters that model the local response of DOMs, bulk ice properties, and neutrino attenuation pull above 2$\sigma$ or hit their boundary.
The same behavior is observed for the parameters that model the neutrino flux, except for two parameters that pull $\sim$$2.3\sigma$: one associated with the cosmic-ray spectrum and one associated with the spectral index that models non-conventional neutrinos below the pivot energy. 
The significance of rejecting the null hypothesis remains consistent across various modeling scenarios for the non-conventional component.
These scenarios include employing a single power law and introducing additional nuisance parameters on the prompt flux and astrophysical $\nu/\bar{\nu}$.
Furthermore, we observe that the preference for a sterile neutrino hypothesis diminishes by a factor of 0.3$\sigma$ when imposing a more stringent constraint of 5\% on the prior width of the overall normalization.

A series of tests were conducted to assess the robustness of the findings.
Consistent outcomes were achieved by conducting independent fits for data collected in different years. 
Additionally, separate fits for the starting and through-going samples revealed a preference for the same sterile neutrino parameter space. 
Furthermore, the results maintained their robustness when fitting different regions in energy and azimuth independently.

Finally, we implemented a test by splitting the data into different zenith regions.
This allowed us to understand the behavior of the fit across different regions of the phase space, where the influence of sterile-like signals varies. 
In fact, vacuum-like oscillations dominate for horizontal events, whereas matter effects prevail for vertically up-going events.
We observed a variation in the significance of rejecting the non-sterile hypothesis for the different fits. 
Specifically, the significance decreased when analyzing only the horizontal events and increased when analyzing only the vertical events.
For example, splitting the samples at $\cos\theta_{\mathrm{reco}}^{\mathrm{v/h}}=-0.2$ resulted in p-values of 0.3\% and 8.2\%, with best-fit points for $(\sin^2(2\theta_{24}),\Delta m^{2}_{41})$ at $(0.2,2.8)$ and $(0.2,5.6)$, respectively. 
Similar trends were identified when splitting the sample at other values of $\cos\theta_{\mathrm{reco}}^{\mathrm{v/h}}$. 
Nevertheless, the preferred regions of all these results remain consistent with those drawn from the full-zenith fit. 

\textbf{Conclusions.---}Studying the atmospheric $\nu_\mu$ spectrum in the TeV regime has evolved into a well-established method for probing sterile neutrinos. 
This work analyzes 10.7 years of data collected with the IceCube experiment, incorporating significant enhancements in event selection, reconstruction, and systematic uncertainty treatment compared to earlier studies.
The data is consistent with the absence of a sterile neutrino with a 3.1\% probability and agrees with previous IceCube analyses.
Notably, this analysis introduces for the first time the examination of two distinct event morphologies, namely starting and through-going events, yielding mutually compatible results.

This result underscores the key role of the IceCube experiment in the context of the 3+1 neutrino landscape.
Consequently, exploring the high-energy muon disappearance channel with other operational neutrino telescopes like KM3NeT~\cite{KM3Net:2016zxf} or Baikal-GVD~\cite{Allakhverdyan:2021vkk} is imperative. 
Moreover, efforts to improve our understanding of the detector and investigate various energy regimes are instrumental for future sterile neutrino analyses. Configurations such as the IceCube Upgrade~\cite{Ishihara:2019aao} and IceCube-Gen2~\cite{IceCube-Gen2:2020qha} are particularly noteworthy in this regard.

\textbf{Acknowledgements.---}The IceCube collaboration acknowledges the significant contributions to this manuscript from the Harvard University, Massachusetts Institute of Technology, and University of Texas at Arlington groups.
We acknowledge the support from the following agencies: USA {\textendash} U.S. National Science Foundation-Office of Polar Programs,
U.S. National Science Foundation-Physics Division,
U.S. National Science Foundation-EPSCoR,
U.S. National Science Foundation-Office of Advanced Cyberinfrastructure,
Wisconsin Alumni Research Foundation,
Center for High Throughput Computing (CHTC) at the University of Wisconsin{\textendash}Madison,
Open Science Grid (OSG),
Partnership to Advance Throughput Computing (PATh),
Advanced Cyberinfrastructure Coordination Ecosystem: Services {\&} Support (ACCESS),
Frontera computing project at the Texas Advanced Computing Center,
U.S. Department of Energy-National Energy Research Scientific Computing Center,
Particle astrophysics research computing center at the University of Maryland,
Institute for Cyber-Enabled Research at Michigan State University,
Astroparticle physics computational facility at Marquette University,
NVIDIA Corporation,
and Google Cloud Platform;
Belgium {\textendash} Funds for Scientific Research (FRS-FNRS and FWO),
FWO Odysseus and Big Science programmes,
and Belgian Federal Science Policy Office (Belspo);
Germany {\textendash} Bundesministerium f{\"u}r Bildung und Forschung (BMBF),
Deutsche Forschungsgemeinschaft (DFG),
Helmholtz Alliance for Astroparticle Physics (HAP),
Initiative and Networking Fund of the Helmholtz Association,
Deutsches Elektronen Synchrotron (DESY),
and High Performance Computing cluster of the RWTH Aachen;
Sweden {\textendash} Swedish Research Council,
Swedish Polar Research Secretariat,
Swedish National Infrastructure for Computing (SNIC),
and Knut and Alice Wallenberg Foundation;
European Union {\textendash} EGI Advanced Computing for research, and Horizon 2020 Marie Sk\l{}odowska-Curie Actions;
Australia {\textendash} Australian Research Council;
Canada {\textendash} Natural Sciences and Engineering Research Council of Canada,
Calcul Qu{\'e}bec, Compute Ontario, Canada Foundation for Innovation, WestGrid, and Digital Research Alliance of Canada;
Denmark {\textendash} Villum Fonden, Carlsberg Foundation, and European Commission;
New Zealand {\textendash} Marsden Fund;
Japan {\textendash} Japan Society for Promotion of Science (JSPS)
and Institute for Global Prominent Research (IGPR) of Chiba University;
Korea {\textendash} National Research Foundation of Korea (NRF);
Switzerland {\textendash} Swiss National Science Foundation (SNSF).

\bibliographystyle{apsrev4-1}
\bibliography{bibliography}

\end{document}